\documentclass[10pt]{article}
\usepackage[dvips]{graphicx}
\usepackage{tabularx}
\usepackage{amsmath,amssymb}
\usepackage{bm}
\newcommand{\eq}{\begin{equation}}
\newcommand{\eqend}{\end{equation}}
\newcommand{\ovl}{\overline}
\newcommand{\A}{\alpha}
\newcommand{\B}{\beta}
\newcommand{\Slash}[1]{\ooalign{\hfill/\hfill\cr\cr$#1$}}
 \addtocounter{section}{0}
 \makeatletter \@addtoreset{equation}{section}
 \makeatother

\title{Supersymmetric Dirac Operator on the Noncommutative Geometry}
\author{Satoshi Ishihara, 
  \footnote{E-mail:satoshi@yukawa.kyoto-u.ac.jp}
\and
Hironobu Kataoka,
  \footnote{E-mail:s499756@hyogo-c.ed.jp}
\and
Atsuko Matsukawa,
  \footnote{E-mail:Atsuko Matsukawa@cap.ocn.ne.jp}
\and 
Hikaru Sato,
  \footnote{E-mail:hikaru\underline{ }sato@gakushikai.jp} \\
{\it Department of Physics, Hyogo University of Education} \\
{\it Shimokume, Kato-shi, Hyogo 673-1494, Japan} \\
\\ 
Masafumi Shimojo \footnote{E-mail:shimo0@ei.fukui-nct.ac.jp} \\
{\it Department of Electronics and Information Engineering, }\\
{\it Fukui National College of Technology,} \\
{\it Geshicho, Sabae-Shi, Fukui 916-8507, Japan}
}
\begin{document}
\maketitle
\begin{abstract}
We extend naturally the spectral triple which define noncommutative geometry (NCG) 
in order to incorporate supersymmetry and obtain supersymmetric Dirac operator 
$\mathcal{D}_M$ which acts on Minkowskian manifold. 
Inversely, 
we can consider the projection which reducts
$\mathcal{D}_M$ to $\Slash{D}$, the Dirac operator of the original spectral triple. 
We investigate properties of the Dirac operator, some of which are inherited from the original 
Dirac operator. $Z/2$ grading and real structure are also supersymmetrically extended. 
Using supersymmetric invariant product, the kinematic terms of chiral and antichiral supermultiplets 
which represent the wave functions of matter particles and their superpartners are provided by $\mathcal{D}_M$. 
Considering the fluctuation given by elements of the algebra 
to the extended Dirac operator, we can expect to obtain vector supermultiplet which includes gauge 
field and to obtain super Yang-Mills theory according to the supersymmetric version of spectral 
action principle.  
\end{abstract}
%
%
\section{Introduction} 

\ \ \, 
The new boson with the mass about 125 GeV/c$^2$ discovered by the LHC experiments turned out to be the long sought Higgs boson of the electroweak theory\cite{ATLAS,CMS,cernpress}.   
In the Lagrangian of gauge theory, the gauge principle gives the term which describes
each gauge boson of $SU(3)$, $SU(2)$, $U(1)$ internal symmetry, while in the standard model(SM)
the Higgs mechanism fixes the terms 
which describe weak bosons and photon in the spontaneaously symmetry
broken theory which is independent of the principle. 
So, the SM has many free parameters and its predictve power is spoiled.

The noncommutative geometry (NCG) was applied to construct the standard model of the elementary particles 
by many authors\cite{rf:Grcoupled,connes1,connes2,Conceptual}.  A remarkable feature of this geometric approach is that both the gauge field and the Higgs field are introduced by internal fluctuations of the metric of NCG. 
The advantage of this approach is that the Higgs field emerges on the same footing as the gauge field and the Higgs coupling constants are related to the gauge coupling constants.

The NCG standard model, extended to include neutrino masses, was constructed\cite{connes8,JHigh}.  Although the standard model provides a remarkably successful description of presently known phenomena, there are some unanswered questions, which suggest the existence of new physics beyond the standard model:  The gauge couplings unification in the renormalization group equation is not viable phenomenologically in the framework of the minimal standard model.  There is also the infamous "hierarchy problem", which is that the Higgs squared mass parameter $m^2_H$ receives enormous quantum corrections from the virtual effects of every particle which couples to the Higgs field\cite{martin}.  In addition, to answer the questions of the origin of the dark matter discovered by the astronomical observation is outside the standard model\cite{Olive}.

It is known that these shortcomings may be remedied by introducing supersymmetry into the standard model.  In order for the NCG standard model to be phenomenologically viable, it is quite desirable to incorporate supersymmetry in the model. 
The purpose of this paper is to investigate how to introduce supersymmetry into NCG and to derive the supersymmetric Dirac operator on the Riemannian manifold whose fluctuation induces vector supermultiplet, 
while we will leave the discussion on the part of supersymmetric Dirac operator 
whose fluctuation induces Higgs supermultiplet to our future papers. 
In section 2, we 
extend the Hilbert space $L^2(M,S)$ to $\mathcal{H}_M$ which includes not only spinor wave function but also its superpartner and 
auxiliary field. It constitutes of the space of chiral supermultiplet and that of antichiral supermultiplet. Then we obtain 
algebra which acts on $\mathcal{H}_M$. We also look for Dirac operator $\mathcal{D}_M$ which acts on $\mathcal{H}_M$ and 
verify that it is supersymmetric and gives the kinetic term of chiral and antichiral supermultiplet 
of matter particles and its superpartners. 
In section 3, we investigate properties of the supersymmetric Dirac operator. 
We can extend the element of the spectral triple to that of the extended triple and vice versa so that 
the supersymmetric Dirac operator has inherited some properties from that of the spectral triple. 
$Z/2$ grading operator and antilinear operator of real structure are also extended to their supersymmetric version.  
In section 4, we describe the conclusions and a discussion on how to introduce vector supermultiplet as internal 
fluctuation of the Dirac operator. 
\section{Supersymmetry and Dirac operator} 

\ \ \, 
The basic element of NCG consists of an involutive algebra $\mathcal{A}$ of operators in a Hilbert space $\mathcal{H}$ and of a self-adjoint unbounded operator $\mathcal{D}$ on $\mathcal{H}$ with compact resolvent such that the commutator $[\mathcal{D}, a]$ is bounded for all elements in $\mathcal{A}$.   
A set of $(\mathcal{A}, \mathcal{H}, \mathcal{D})$  
 is named a spectral triple or a $K$-cycle\cite{connes0}.
 
The geometry of a compact Riemannian manifold $M$ is described by the {\it canonical} spectral triple $(C^\infty(M), L^2(M,S), \Slash{D})$, where the Hilbert space $L^2(M,S)$ consists of square-integrable sections of the spinor bundle $S \rightarrow M$ and the operator $\Slash{D}$ is the Dirac operator on the spinor bundle. For the flat space-time manifold, $M$ is the four dimensional Euclidean manifold with the signature $\eta^{\mu\nu} = (1,1,1,1)$.  After the calculation of the spectral action and other quantities in NCG in the Euclidean signature, we translate the results into the physical ones in the Minkowskian space-time manifold with the signature $g^{\mu\nu} = (-1,1,1,1)$ by the Wick rotation, $t \rightarrow it$.

In order to introduce supersymmetry, we must work in the Minkowskian space-time $M$. So, 
let us construct $(\mathcal{A}_M, \mathcal{H}_M, \mathcal{D}_M)$ on the Minkowskian manifold $M$ in the following way:
Since supersymmetry is expressed on the equal number of fermionic and bosonic degrees of freedom, $\mathcal{H}_M$ consists of spinor and scalar $L^2$ functions. The spinorial subspace of $\mathcal{H}_M$, which is obtained by the projection $P \mathcal{H}_M$, corresponds to $L^2(M,S)$ of the spectral triple if we work in the Euclidean signature. Accordingly, the projected algebra $P \mathcal{A}_M P$ corresponds to $C^\infty(M)$ and $P \mathcal{D}_M P$ to the Dirac operator $\Slash{D}$. Hereafter, we will call $(\mathcal{A}_M, \mathcal{H}_M, \mathcal{D}_M)$ simply "the triple".        

Now let us construct $\mathcal{H}_M$ concretely. A Dirac spinor $\psi(x)$
 is expressed by two Weyl spinors $\psi_+(x)$ and $\psi_-(x)$ as 
\eq
\psi = \begin{pmatrix}
\psi_+ \\
\psi_-
\end{pmatrix}
\eqend
in the Weyl basis. Here $\psi_+(x)$ transforms as the $(\frac{1}{2}, 0)$ representation of the Lorentz group, $SL(2,C)$, and is denoted by $\psi_{+\A}(x)$ with undotted indices $\A, \A = 1,2$. Let us call  $\psi_+(x)$ chiral spinor. Similarly,  $\psi_-(x)$ transforms as the  $(0, \frac{1}{2})$ representation of $SL(2,C)$, which is denoted by $\psi_-^{\dot{\A}}(x)$ with dotted indices $\dot{\A}, \dot{\A} = 1,2$, and is 
called antichiral spinor. The indices $\A$ are raised and lowered with the antisymmetric tensors $\varepsilon^{\A\B}$ and $\varepsilon_{\A\B}$, where $\varepsilon^{12} = \varepsilon_{21} = 1$. The same holds for dotted indices $\dot{\A}$. 
The $\varepsilon$-tensor is also used to raise the indices of the $\sigma$-matrices as $\bar{\sigma}^{\mu\dot{\A}\A} = \varepsilon^{\dot{\A}\dot{\B}} \varepsilon^{\A\B} \sigma^\mu_{\B\dot{\B}}$. 
Here and in what follows we use the notation and convention of Wess-Bagger\cite{WessBagger}.

In order to introduce supersymmetry we need two complex scalar functions $\varphi_+(x)$ and $F_+(x)$ which are superpartners of a chiral spinor $\psi_{+\A}(x)$. Here $\varphi_+(x)$ and $ F_+(x)$ have mass dimension 1 and 2, respectively, and $\psi_{+\A}(x)$ have mass dimension $\frac{3}{2}$. These functions are supposed to obey the following supersymmetry transformation:
\eq
\left\{
\begin{array}{lcl}
\delta_\xi \varphi_+ &=& \sqrt{2} \xi^\A \psi_{+\A}
\\
\delta_\xi \psi_{+\A} &=& i \sqrt{2} \sigma^\mu_{\A\dot{\A}} \bar{\xi}^{\dot{\A}} \partial_\mu \varphi_+ + \sqrt{2} \xi_\A F_+
\\
\delta_\xi F_+  &=& i \sqrt{2} \bar{\xi}_{\dot{\A}} \bar{\sigma}^{\mu\dot{\A}\A} \partial_\mu \psi_{+\A}.
\end{array}
\right. 
\label{eq2.2}
\eqend
The set of functions
\eq
(\Psi_+)_i = \left( \varphi_+, \psi_{+\A}, F_+  \right)^T, \ \ i = 1, 2, 3
\eqend
is called a chiral supermultiplet, which is denoted here in the vector notation.
For the antichiral spinor $\psi_-^{\dot{\A}}(x)$, we introduce $\varphi_-(x)$ and $F_-(x)$ which obey the supersymmetry transformation given by
\eq
\left\{
\begin{array}{lcl}
\delta_{{\xi}} \varphi_- &=& \sqrt{2} \bar{\xi}_{\dot{\A}} \, {\psi}_-^{\dot{\A}} \\ 
\delta_{{\xi}} {{\psi}}^{\dot{\A}}_- &=& i \sqrt{2}  \ovl{\sigma}^{\mu\dot{\A}\A} \,{\xi}_\A \partial_\mu \varphi_- + \sqrt{2}\ \bar{\xi}^{\dot{\A}} F_- , \\
\delta_{{\xi}} F_- &=& i \sqrt{2} \,{\xi}^\A \sigma^\mu_{\A\dot{\A}} \partial_\mu {\psi}_-^{\dot{\A}}
\end{array} \right.
\label{eq2.4}
\eqend
and the set of functions
\eq
(\Psi_-)_{\bar{i}} = \left( \varphi_-, \psi_-^{\dot{\A}}, F_-  \right)^T, \ \ \bar{i} = 1, 2, 3
\eqend
is called an antichiral supermultiplet. 
Then $\mathcal{H}_M$ is the direct sum of two subsets, $\mathcal{H}_+$ 
and $\mathcal{H}_-$:
\begin{equation}
\mathcal{H}_M = \mathcal{H}_+\oplus \mathcal{H}_-,
\end{equation}
where $\mathcal{H}_+ $ is the space of chiral supermultiplets and $\mathcal{H}_-$ is the space of antichiral supermultiplets 
and the element of $\mathcal{H}_M$ is given by
\begin{equation}
\Psi= 
\begin{pmatrix}
(\Psi_+)_i \\
(\Psi_-)_{\bar{i}}
\end{pmatrix} \in \mathcal{H}_M.
\label{eq2.6}
\end{equation}

A supersymmetric invariant product of wave functions $\Psi $, $\Psi^\prime $ is defined by
\begin{equation}
(\Psi^\prime,\Psi) = \int_M \Psi^{\prime\dagger} {\rm \Gamma}_0\Psi d^4x, \label{superinvproduct}
\end{equation}
where
\begin{equation}
\Gamma_0 = \begin{pmatrix}
0 & \mathit{\Gamma}_0\\
\mathit{\Gamma}_0 & 0
\end{pmatrix},\ 
\mathit{\Gamma}_0 =\begin{pmatrix}
0 & 0 & 1\\
0 & -{\bf 1}_{2\times 2} & 0\\
1 & 0 & 0
\end{pmatrix}.
\end{equation}

Now let us construct an algebra $\mathcal{A}_M$ of operators in $\mathcal{H}_M$. For the basis of $\mathcal{H}_M$ given by Eq.(\ref{eq2.6}), the element of $\mathcal{A}_M$ is expressed by the following matrix form:
\eq
a_{i\bar{i};j\bar{j}} = \begin{pmatrix}
u_{ij} & 0 \\
0 & \bar{v}_{\bar{i}\,\bar{j}}
\end{pmatrix},
\label{eq2.7}
\eqend 
where $u_{ij}$ is given by the following triangular matrix,
\eq
u_{ij} = \frac{1}{m_0} \begin{pmatrix}
\varphi_\eta  & 0 & 0 \\
\eta_{\A} & \varphi_\eta  & 0 \\
F_{\eta} & -\eta^\A & \varphi_\eta 
\end{pmatrix}. 
\label{eq2.8}
\eqend
Here the scalar functions $\varphi_\eta(x), F_\eta(x)$ and the spinor functions $\eta_{\A}(x)$ are supposed to form a chiral supermultiplet $(\varphi_\eta, \eta_\A, F_\eta)$ and obey the same type of the supersymmetry transformation given by Eq.(\ref{eq2.2}). In Eq.(\ref{eq2.8}),  $m_0$ stands for the mass parameter which was inserted to adjust the mass dimension. 
Since $(\varphi_\eta, \eta_\A, F_\eta)$ is the chiral supermultiplet, the triangular matrix $u_{ij}$ given by Eq.(\ref{eq2.8}) obeys the following multiplication rule:
\begin{align}
& \hspace{1.8mm} (u_3)_{ik} =  (u_1)_{ij} (u_2)_{jk},
\label{eq2.9} \\
&\left\{
\begin{array}{lcl}
\varphi_{\eta 3} &=& \varphi_{\eta 1} \,\varphi_{\eta 2} /m_0, \\
\eta_{3 \A} &=& (\eta_{1\A}\, \varphi_{\eta 2} + \varphi_{\eta 1} \,\eta_{2\A})/m_0, \\
F_{\eta_3} &=& (\varphi_{\eta 1}\, F_{\eta_2} + F_{\eta_1} \,\varphi_{\eta 2} - \eta^\A_1 \,\eta_{2\A})/m_0.
\end{array}
\right.
\label{eq2.10}
\end{align}
As a matter of fact, $\varphi_{\eta 3}, \eta_{3\A}$ and $F_{\eta_3}$ 
transform again as Eq.(\ref{eq2.2}) and form a chiral supermultiplet.

For the antichiral sector, $\bar{v}_{\bar{i}\,\bar{j}}$ in Eq.(\ref{eq2.7}) is given by
\eq
(\bar{v})_{\bar{i}\,\bar{j}} = \frac{1}{m_0} \begin{pmatrix}
\varphi_\chi^* & 0 & 0 \\
{\bar{\chi}}^{\dot{\A}} & \varphi_\chi^* & 0 \\
F_{\chi}^* & -\bar{\chi}_{\dot{\A}} & \varphi_\chi^*
\end{pmatrix},  
\label{eq2.11}
\eqend
where $\varphi_\chi^*(x), F^*_\chi(x)$ and $\bar{\chi}^{\dot{\A}}(x)$ are chosen to form an antichiral supermultiplet and they obey the supersymmetry transformation given by Eq.(\ref{eq2.4}). The multiplication rule for $\bar{v}_{\bar{i}\,\bar{j}}$ is given by
\begin{align}
& \hspace{1.8mm} (\bar{v}_3)_{\bar{i}\,\bar{k}} =  (\bar{v}_1)_{\bar{i}\,\bar{j}} (\bar{v}_2)_{\bar{j}\,\bar{k}},
\label{eq2.12} \\
&\left\{
\begin{array}{lcl}
\varphi^*_{\chi 3} &=& \varphi^*_{\chi 1} \,\varphi^*_{\chi 2} /m_0, \\
\bar{\chi}^{\dot{\A}}_{3} &=& (\bar{\chi}^{\dot{\A}}_{1} \,\varphi^*_{\chi 2} + \varphi^*_{\chi 1} \,\bar{\chi}^{\dot{\A}}_{2})/m_0, \\
F^*_{\chi_3} &=& (\varphi^*_{\chi 1} \,F^*_{\chi_2} + F^*_{\chi_1} \,\varphi^*_{\chi 2} - \bar{\chi}_{1\dot{\A}} \,\bar{\chi}_2^{\dot{\A}})/m_0.
\end{array}
\right.
\label{eq2.13}
\end{align}
Taking into account Eq.(\ref{eq2.9}) and Eq.(\ref{eq2.12}), we obtain the multiplication rule for the element of $\mathcal{A}_M$ given by Eq.(\ref{eq2.7}) as
\eq
(a_3)_{i\bar{i};k\bar{k}} = \sum_{j\bar{j}}(a_1)_{i\bar{i};j\bar{j}} (a_2)_{j\bar{j};k\bar{k}}.
\eqend

The multiplication rule of Eq.(\ref{eq2.9}) with Eq.(\ref{eq2.10}) is easily understood by the help of the superfield notation. In order to do this, we introduce  anticommuting parameters $\theta^\A, \bar{\theta}^{\dot{\A}}$. Then $u_{ij}$ of Eq.(\ref{eq2.8}) can be expressed by the following chiral superfield,
\eq
U(x_+) = \frac{1}{m_0} \left\{\varphi_\eta(x_+) + \sqrt{2} \theta^\A \eta_\A(x_+) + \theta^\A\theta_\A F_\eta(x_+)\right\}, 
\label{}
\eqend
where $x_{\pm}^\mu = x^\mu \pm \theta^\A \sigma^\mu_{\A\dot{\A}} \bar{\theta}^{\dot{\A}}$. Since the product of two chiral superfields is again a chiral superfield such that $U_3 = U_1 U_2$, we can deduce the relation given by Eq.(\ref{eq2.10}).

For the antichiral supermultiplet expressed by Eq.(\ref{eq2.11}), the corresponding antichiral superfield is given by
\eq
\bar{V}(x_-) =  \frac{1}{m_0} \left\{\varphi_\chi^*(x_-) + \sqrt{2}\bar{\theta}_{\dot{\A}} \bar{\chi}^{\dot{\A}}(x_-) + \bar{\theta}_{\dot{\A}}\bar{\theta}^{\dot{\A}} F^*_{\chi}(x_-)  \right\}.
\eqend
The multiplication formula for the antichiral supermultiplets given by Eq.(\ref{eq2.13}) is again obtained from 
$\bar{V}_3 = \bar{V}_1 \bar{V}_2$. 

From Eq.(\ref{eq2.7}), we see that the element of $\mathcal{A}_M$ 
is expressed by
\begin{equation}
a_{i\bar{i};j\bar{j}}=a_{ij}+\bar{a}_{\bar{i}\bar{j}},
\end{equation} 
and 
\begin{equation}
a_{ij}=\begin{pmatrix}
u_{ij} & 0\\
0 & 0
\end{pmatrix} \in \mathcal{A}_+,\ \ 
\bar{a}_{\bar{i}\bar{j}} =\begin{pmatrix}
0 & 0\\
0 & \bar{v}_{\bar{i}\bar{j}}
\end{pmatrix}\in \mathcal{A}_-,
\end{equation}
where $\mathcal{A}_+$ is the subspace which acts on $\mathcal{H}_+$ and $\mathcal{A}_-$ is the subspace which acts on $\mathcal{H}_-$ 
so that $\mathcal{A}_M$ is the direct sum of $\mathcal{A}_+$ and $\mathcal{A}_-$:
\begin{equation}
\mathcal{A}_M = \mathcal{A}_+ \oplus \mathcal{A}_-.
\end{equation}

In the rest of this section we construct the supersymmetrically extended Dirac operator $\mathcal{D}_M$ in $\mathcal{H}_M$, which is supersymmetric and reduces to the usual Dirac operator $\Slash{D}$ of the noncommutative geometry if $\mathcal{H}_M$ is restricted to the spinorial subspace $L^2(M,S)$. For the basis of $\mathcal{H}_M$ expressed by Eq.(\ref{eq2.6}), the $\gamma$ matrices in the Weyl representation amount to
\eq
\gamma^\mu = 
\left(
\begin{array}{@{\,}ccc|ccc@{\,}}
 & & & 0 & {\bf{0}}_{1\times 2} & 0 \\ 
 & {\bf{0}}_{4\times 4} & & {\bf{0}}_{2\times 1} & \sigma^\mu & {\bf{0}}_{2\times 1} \\
 & & & 0 & {\bf{0}}_{1\times 2} & 0 \\
 \hline 
0 & {\bf{0}}_{1\times 2} & 0 & & & \\
{\bf{0}}_{2\times 1} & \bar{\sigma}^\mu & {\bf{0}}_{2\times 1} & & {\bf{0}}_{4\times 4} & \\
0 & {\bf{0}}_{1\times 2} & 0 & & &
\end{array} \right),
\eqend 
where ${\bf{0}}_{n\times m}$ stands for the $n\times m$ null matrix.
The Dirac operator in $\mathcal{H}_M$ is now given by
\eq
\Slash{D} = i\, \gamma^\mu \partial_\mu
\eqend
Although $\Slash{D}$ is not supersymmetric, we can show that the modified operator $\mathcal{D}_M$ defined by
\eq
i\mathcal{D}_M = \Slash{D} + \Delta, \label{DM}
\eqend
with
\eq
\Delta = \left(
\begin{array}{@{\,}ccc|ccc@{\,}}
 & & & 0 & {\bf{0}}_{1\times 2} & 1 \\ 
 & {\bf{0}}_{4\times 4} & & {\bf{0}}_{2\times 1} & {\bf{0}}_{2\times 2} & {\bf{0}}_{2\times 1} \\
 & & & \square & {\bf{0}}_{1\times 2} & 0 \\
 \hline 
0 & {\bf{0}}_{1\times 2} & 1 & & & \\
{\bf{0}}_{2\times 1} & {\bf{0}}_{2\times 2} & {\bf{0}}_{2\times 1} & & {\bf{0}}_{4\times 4} & \\
\square & {\bf{0}}_{1\times 2} & 0 & & &
\end{array} \right),
\eqend   
and $\square = \partial^\mu \partial_\mu$ is supersymmetric.

In order to verify that $i\mathcal{D}_M$ is supersymmetric, let us express the supersymmetry transformation $\delta_\xi$ given by Eq.(\ref{eq2.2}) and Eq.(\ref{eq2.4}) on the basis of $\mathcal{H}_M$ as follows: 
\eq
\delta_\xi = \sqrt{2} \left(
\begin{array}{@{\,}ccc|ccc@{\,}}
0 & \xi^\A  & 0 &  &  &  \\
i \sigma_{\A\dot{\A}}^\mu \bar{\xi}^{\dot{\A}} \partial_\mu & {\bf{0}}_{2\times 2} &  \xi_\A &  & {\bf{0}}_{4\times 4} &  \\
0 & i \bar{\xi}_{\dot{\A}} \bar{\sigma}^{\mu\dot{\A}\A} \partial_\mu & 0 &  &  &  
\vspace{0.5mm} \\
\hline
 &  &  & 0 & \ovl{\xi}_{\dot{\A}} & 0 \\
 & {\bf{0}}_{4\times 4} &  & i \bar{\sigma}^{\mu \dot{\A}\A} {\xi}_{{\A}} \partial_\mu & {\bf{0}}_{2\times 2} & \bar{\xi}^{\dot{\A}} \\
 &  &  & 0 & i {\xi}^{{\A}} {\sigma}^{\mu}_{\A\dot{\A}} \partial_\mu & 0
\end{array}
\right).
\eqend
Then, we can show that
\eq
\left[\, \delta_\xi, \mathcal{D}_M \right] = 0
\eqend
In what follows we shall call $\mathcal{D}_M$ the "supersymmetrically extended Dirac operator" or simply "extended Dirac operator".  
$\mathcal{D}_M$ is written in the matrix form as follows:
\begin{equation}
\mathcal{D}_M = -i\begin{pmatrix}
0 & \bar{\mathcal{D}}_{i\bar{j}} \\
\mathcal{D}_{\bar{i}j} & 0
\end{pmatrix}, 
\label{DM0}
\end{equation}
where
\begin{equation}
\mathcal{D}_{\bar{i}j}=\begin{pmatrix}
0 & 0 & 1\\
0 & i\bar{\sigma}^\mu\partial_\mu & 0\\
\Box & 0 & 0
\end{pmatrix},\ 
\bar{\mathcal{D}}_{i\bar{j}}= \begin{pmatrix}
0 & 0 & 1\\
0 & i\sigma^\mu\partial_\mu & 0\\
\Box & 0 & 0
\end{pmatrix}.
\label{DM01}
\end{equation}
The operator $\mathcal{D}_{\bar{i}{j}}$ operates on $(\Psi_+)_j$ and generates the element 
in $\mathcal{H}_-$ in the following way,
\begin{align}
&\hspace{5mm} (\Psi'_-)_{\bar{i}} = \frac{1}{m_0} \mathcal{D}_{\bar{i}{j}} (\Psi_+)_j,
\label{PsiPrimem} 
\\
&\left\{
\begin{array}{lcl}
\varphi'_-(x) = F_+(x)/m_0, \\
{\psi}'^{\dot{\A}}_-(x) = i \ovl{\sigma}^{\mu\dot{\A}\A} \partial_\mu \psi_{+\A} (x) /m_0, \\
F'_-(x) = \square \varphi_+(x)/m_0.
\end{array}
\right.
\label{eq2.14}
\end{align}
As a matter of fact, we can confirm that the left-hand side of Eq.(\ref{eq2.14}) obeys  Eq.(\ref{eq2.4}) 
which is the transformation law 
for the antichiral supermultiplet by  applying Eq.(\ref{eq2.2}) on the right-hand side.  Similarly, for the operator $\ovl{\mathcal{D}}_{i \bar{j}}$ we have the following formula:
\begin{align}
&\hspace{5mm} (\Psi'_+)_{{i}} = \frac{1}{m_0} \ovl{\mathcal{D}}_{{i}\bar{j}} (\Psi_-)_{\bar{j}},
\label{eq2.15} 
\end{align}
then Eq.(\ref{PsiPrimem}) and Eq.(\ref{eq2.15}) are put together in the following form:
\eq
\Psi' = \frac{i}{m_0} \mathcal{D}_M \Psi .
\label{eq2.17}
\eqend

Using the definition of supersymmetric invariant product (\ref{superinvproduct}), 
the kinetic term in the action constructed by chiral and antichiral supermultiplets of matter fields with 
their superpartners is  
expressed by 
\begin{align}
I_{kinetic} = (\Psi,i\mathcal{D}_M\Psi) & = 
\int_M d^4x
\left(
\varphi_+^\ast\Box \varphi_+ - i\bar{\psi}_+\bar{\sigma}^\mu\partial_\mu\psi_+ +F_+^\ast F_+ 
+ \varphi_- \Box \varphi_-^\ast -i\psi_-\sigma^\mu\partial_\mu \bar{\psi}_-+F_-F_-^\ast
\right) \label{Ikinetic}
\end{align}  
\section{Noncommutative geometry and Supersymmetry} 
\ \ \ \ \  
Since models based on NCG in the flat space-time is constructed in the Euclidean space-time, 
if we see the correspondence between "the triple" and the spectral triple which defines a NCG, 
we must transform the variables in Minkowskian coordinates to Euclidean ones. 
The space-time variables are transformed by Wick rotation as follows:
\begin{equation}
x^0 \rightarrow -ix^0.  
\end{equation}  
The algebra of $SL(2,C)$ turns out the algebra of $SU(2)\otimes SU(2)$ under the 
rotation. 
The Weyl spinors which transform as $(\frac{1}{2},0)$, $(0,\frac{1}{2})$ of $SL(2,C)$ 
are to be replaced by $(\frac{1}{2},0)$ and $(0,\frac{1}{2})$ representations of 
$SU(2)\otimes SU(2)$,respectively. The spinors which have appeared in $\mathcal{H}_M$ 
are replaced as follows:
\begin{align}
\psi_{+\alpha} \rightarrow \rho_{\alpha}, 
& \ \psi_{+}^\alpha \rightarrow \rho^{\alpha}, \label{replacepsi+}\\
\psi_{-}^{\dot{\alpha}}\rightarrow \omega^{\dot{\alpha}}, & \ 
\psi_{-\dot{\alpha}} \rightarrow \omega_{\dot{\alpha}},\label{replacepsi-}
\end{align}
where spinors with indices $\alpha$ transform as $(\frac{1}{2},0)$ and those with indices $\dot{\alpha}$ 
transform as $(0,\frac{1}{2})$ of $SU(2)\otimes SU(2) $,respectively.  
The upper index is related to the complex conjugate of the lower 
index by $\rho^1=\rho_2^\ast$, $\rho^2=-\rho_1^\ast$, $\omega^{\dot{1}}=\omega_{\dot{2}}^\ast$, 
$\omega^{\dot{2}}=-\omega_{\dot{1}}^\ast $. 
The metric and Pauli matrices which have appeared in the extended Dirac operator are to be replaced by 
\begin{align}
g^{\mu\nu} = (-1,1,1,1) & \rightarrow \eta^{\mu\nu}=(1,1,1,1) \label{replacemetric}\\
\sigma^\mu  \rightarrow \sigma^\mu_E = (\sigma^0,i\sigma^i), & \ \  
\bar{\sigma}^\mu \rightarrow \bar{\sigma}_E^\mu =(\sigma^0,-i\sigma^i).\label{replacesigma}
\end{align}

Embedding these expressions (\ref{replacepsi+})$\sim$ (\ref{replacesigma}), the triple is  
rewritten in the Euclidean signature.
%
%
The basis of $\mathcal{H}_M$ is denoted by the same form as (\ref{eq2.6}), 
but now $\Psi_+$ and $\Psi_-$  are given by 
\begin{align}
(\Psi_+)_i & =(\varphi_+,\rho_\alpha,F_+)^T, \label{PsiE+} \\
(\Psi_-)_{\bar{i}} & =(\varphi_- ,\omega^{\dot{\alpha}},F_-)^T. \label{PsiE-}
\end{align}

The elements of $\mathcal{A}_M$  which correspond to (\ref{eq2.8}) and (\ref{eq2.11}) is now 
given by 
\begin{align}
(u_a)_{ij} & = \frac{1}{m_0}
\begin{pmatrix}
\varphi_{\eta a} & 0 & 0\\
\eta_{Ea\alpha} & \varphi_{\eta a} & 0 \\
F_{\eta a} & -\eta_{Ea}^\alpha & \varphi_{\eta a}
\end{pmatrix} \in \mathcal{A}_+, 
\\
(\bar{v}_a)_{\bar{i}\bar{j}} & = \frac{1}{m_0}
\begin{pmatrix}
\varphi_{\chi a}^\ast & 0 & 0\\
\chi_{E a}^{\dot{\alpha}} & \varphi_{\chi a}^\ast & 0 \\
F_{\chi a}^\ast & -\chi_{E a\dot{\alpha}}^\alpha & \varphi_{\chi a}^\ast
\end{pmatrix} \in \mathcal{A}_-,
\end{align}
where $\eta_{Ea\alpha}$, $\chi_{Ea}^{\dot{\alpha}}$ are $(\frac{1}{2},0)$ and 
$(0,\frac{1}{2})$ of $SU(2)\otimes SU(2)$,respectively. 
For the extended Dirac operator on the Minkowskian manifold in 
(\ref{DM0}), the transformed form is given by
\begin{equation}
\mathcal{D}_M = -i\begin{pmatrix}
0 & \bar{\mathcal{D}}_E \\
\mathcal{D}_E & 0
\end{pmatrix}, \label{DME0}
\end{equation}
where
\begin{equation}
\mathcal{D}_{E \bar{i}j}=\begin{pmatrix}
0 & 0 & 1\\
0 & \bar{\sigma}_E^\mu\partial_\mu & 0\\
\Box_E & 0 & 0
\end{pmatrix},\ 
\bar{\mathcal{D}}_{E i\bar{j}}= \begin{pmatrix}
0 & 0 & 1\\
0 & \sigma_E^\mu\partial_\mu & 0\\
\Box_E & 0 & 0
\end{pmatrix}, \label{DME}
\end{equation}
with 
\begin{equation}
\Box_E = \eta^{\mu\nu}\partial_\mu\partial_\nu = \partial_0^2+\partial_i^2.
\end{equation}
The invariant product under Euclidean supersymmetry transformation 
is given by the 
same form as (\ref{superinvproduct}), but $\Gamma_0$ should be replaced by
\begin{equation}
\Gamma_0 =\begin{pmatrix}
\mathit{\Gamma}_0 & 0\\
0 &\mathit{\Gamma}_0
\end{pmatrix}. \label{EGamma}
\end{equation} 

The triple $(\mathcal{A}_M,\mathcal{H}_M,\mathcal{D}_M)$ is supersymmetrically extended from the spectral triple which 
specify a NCG. Inversely, we can consider the operator $P$ which project the triple to the original spectral triple 
embedded in it. On the basis of (\ref{PsiE+}),(\ref{PsiE-}), $P$  is given by
\begin{equation}
P=
\left(\begin{array}{c|c}
\begin{array}{ccc}
0 & 0 & 0 \\
0 & 1_2 & 0 \\
0 & 0 & 0 
\end{array} & \bf{0_{4\times 4}} \\
\hline
\bf{0_{4\times 4}} &
\begin{array}{ccc}
0 & 0 & 0 \\
0 & 1_2 & 0 \\
0 & 0 & 0 
\end{array}
\end{array}\right).
\end{equation}
In fact, the actions of $P$ for the elements of the triple are expressed by 
\begin{equation}
P\Psi = \begin{pmatrix}
0 \\
\psi_+\\
0\\
0\\
\psi_- \\
0
\end{pmatrix},\ \ 
Pa_{i\bar{i};j\bar{j}}P =
\left(
\begin{array}{c|c}
\begin{array}{ccc}
0 & 0 & 0 \\
0 & \varphi_\eta & 0\\
0 & 0 & 0
\end{array} & \bf{0_{4\times 4}}\\
\hline
\bf{0_{4\times 4}} &
\begin{array}{ccc}
0 & 0 & 0 \\
0 & \varphi_\chi^\ast & 0 \\
0 & 0 & 0 
\end{array}
\end{array}
\right),\ \
\begin{array}{c}
Pi\mathcal{D}_MP= \Slash{D}.
\end{array}
\end{equation}

Since $\mathcal{D}_M$ is supersymmetrically extended form $\Slash{D}$, some properties which the inverse of 
line element $ds$ should have due to the axioms of NCG are inherited. \\
\begin{enumerate}
\item[(a)] $\mathcal{D}_M$ has the real eigenvalues.\\
Under the supersymmetric invariant product (\ref{superinvproduct}), the definition of self-adjoint is 
given by
\begin{equation}
(\Psi_1,i\mathcal{D}_M \Psi_2) =(\Psi_2,i\mathcal{D}_M \Psi_1)^\ast.
\end{equation}   
This condition is rewritten to 
\begin{equation}
(\Gamma_0i\mathcal{D}_M)^\dagger = \Gamma_0i\mathcal{D}_M,
\end{equation}
which we can verify easily. Indeed, $\mathcal{D}_M$ has the real eigenvalues. 
We compute the eigenvalues of $\mathcal{D}_M$ for the Euclidean $d$-dimensional torus, $M = T^d$ with circumference $2\pi$, where $d$ is the space-time dimension, $d = 4$. After a Fourier transform, 
\begin{align}
\Psi(x) &= \Psi_n \, e^{-i \sum n_i x_i} \mbox{ } (n_i = 0, \pm 1, \cdots),
\label{eq3.12}
\end{align}
the eigenvalue equation reads as follows,
\eq
\mathcal{D}_M(n) \Psi_n = \lambda_n \Psi_n,
\label{eq3.13}
\eqend
where
\begin{align}
\mathcal{D}_M(n) &= \begin{pmatrix}
0 & \ovl{\mathcal{D}}(n) \\
\mathcal{D}(n) & 0
\end{pmatrix}.
\label{eq3.14}
\end{align}
In Eq.(\ref{eq3.14}),  $\mathcal{D}(n)$ and $\ovl{\mathcal{D}}(n)$ are given by
\eq
\mathcal{D}(n) = \begin{pmatrix}
0 & 0 & 0 & -i \\
0 &  n_0 +i n_3 & i n_1+ n_2 & 0 \\
0 & i n_1 - n_2 &  n_0 -i  n_3 & 0 \\
i n^2 & 0 & 0 & 0
\end{pmatrix},
\label{eq3.15}
\eqend
and
\eq
\ovl{\mathcal{D}}(n) = \begin{pmatrix}
0 & 0 & 0 & -i \\
0 &  n_0 -i n_3 & -i n_1- n_2 & 0 \\
0 & -i n_1 + n_2 &  n_0 + i n_3 & 0 \\
i n^2 & 0 & 0 & 0
\end{pmatrix},
\label{eq3.16}
\eqend
where $n^2 = n_0^2 + n_1^2 + n_2^2 + n_3^2$. The characteristic equation amounts to
\begin{align}
\mbox{det}\left| \mathcal{D}_M(n) - \lambda \mathbf{1}_8 \right| &= (\lambda^2 - n^2)^4
= 0,
\label{eq3.17}
\end{align}
which gives the fourthly degenerate eigenvalues of $\mathcal{D}_M$ as
\eq
\lambda_n = \pm \sqrt{n_0^2 + n_1^2 + n_2^2 + n_3^2}.
\label{eq3.18}
\eqend
For large $|\lambda_n|$, there are about $(\pi^2/2) |\lambda_n|^4$ eigenvalues inside the four dimensional ball with the radius $|\lambda_n|$. If we arrange $|\lambda_n|$ in an increasing sequence, we obtain
\eq
|\lambda_n| \approx \left( \frac{2n}{\pi^2} \right)^{1/4},
\label{eq3.19}
\eqend
for large $n$. So $ds =\mathcal{D}_M^{-1}$ is an infinitesimal of order $1/d$.
\item[(b)] The resolvent $R(\lambda; \mathcal{D}_M) = (\mathcal{D}_M - \lambda\mathbf{1}_8)^{-1}, \lambda \notin \mbox{R}$ of $\mathcal{D}_M$ is compact. 

As a matter of fact, for any $\varepsilon > 0$ with sufficiently large $N$
the norm of the resolvent obeys the following relation:
\begin{align}
||R(\lambda; \mathcal{D}_M)  || &< | \lambda_{N+1} -\lambda |^{-1} 
< \varepsilon,
\label{eq3.21}
\end{align}
on the orthogonal  of a $N$ dimensional subspace of $\mathcal{H}_M$.
\end{enumerate}

On the other hand, since $\mathcal{D}_M$ includes d'Alembertian $\Box$ as an matrix element, $[\mathcal{D}_M,a]$ 
is not unbounded for $a\in \mathcal{A}_M$, while $[\Slash{D},a]$ is bounded for $a\in C^\infty(M)$. Due to this fact, we shall note that the triple extended from 
the spectral triple does not induce new NCG.

In addition to (a) and (b), $Z/2$ grading operator $\gamma_M$ and antilinear operator $J_M$ which 
gives the real structure can be defined on the triple and  
the similar (anti-)commutation relations with $\mathcal{D}_M$ can be obtained. 
$Z/2$ grading $\gamma_M$ is given by an operator in $\mathcal{H}_M$ which is in the Euclidean signature defined by
\begin{equation}
\gamma_M = \begin{pmatrix}
-1 & 0\\
0 & 1
\end{pmatrix}.
\end{equation}
So the anti-commutation relation with $\mathcal{D}_M$ is given by
\begin{equation}
\gamma_M\mathcal{D}_M =-\mathcal{D}_M \gamma_M.
\end{equation}

For the state $\Psi\in \mathcal{H}_M$ the charge conjugate state 
$\Psi^c$ is given by
\begin{align}
\Psi^c &= \begin{pmatrix}
\Psi_+^c \\
\Psi_-^c  
\end{pmatrix} , 
\label{eq3.25}
\end{align}
and
\begin{align}
(\Psi_+^c)_i &= ( \varphi_+^*, \rho^\A,  F_+^*)^T,
\label{eq3.26} \\
(\Psi_-^c)_{\bar{i}} &= ( \varphi_-^*, \omega_{\dot{\A}},  F_-^* )^T.
\label{eq3.27}
\end{align} 
Then let us define the antilinear operator $\mathcal{J}_M$ by
\eq
\Psi^c  = \mathcal{J}_M \Psi = C \Psi^*,
\label{eq3.28}
\eqend
so that it is given by
\eq
\mathcal{J}_M = C \otimes * ,
\label{eq3.29}
\eqend
where $C$ is the following charge conjugation matrix;
\eq
C = \left(
\begin{array}{@{\,}ccc|ccc@{\,}}
1 & & & & & \\
 & \varepsilon^{\A\B} & & & \bm{{0}} & \\ 
 & & 1 &  & & \\ \hline
 & & & 1 & & \\
 & \bm{{0}} & &  & \varepsilon_{\dot{\A}\dot{\B}} & \\
  & & &  & & 1
\end{array}
\right),
\label{eq3.30}
\eqend
and $*$ is the complex conjugation (Hermitian conjugation for matrices). 
The operator $\mathcal{J}_M$  obeys the following relation:
\begin{align}
\mathcal{J}_M \gamma_M &= \gamma_M \mathcal{J}_M.
\label{eq3.31} 
\end{align}

The real structure $J_M$ is now expressed on the basis $(\Psi, \Psi^c)^T$ in the following matrix form:
\eq
J_M = \begin{pmatrix}
0 & \mathcal{J}_M^{-1} \\
\mathcal{J}_M & 0
\end{pmatrix}.
\label{eq3.32}
\eqend
On the same basis, the Dirac operator $D_M$ and the Z/2 grading $\Gamma_M$ is expressed by
\eq
D_M = \begin{pmatrix}
\mathcal{D}_M & 0 \\
0 & \mathcal{J}_M \mathcal{D}_M \mathcal{J}_M^{-1}
\end{pmatrix},
\label{eq3.33}
\eqend
and
\eq
\Gamma_M = \begin{pmatrix}
\gamma_M & 0 \\
0 & \gamma_M
\end{pmatrix}.
\label{eq3.34}
\eqend
The real structure defined by Eq.(\ref{eq3.32}) satisfies the following relations:
\begin{align}
J_M^2 &= 1 ,
\label{eq3.35} \\
J_M {D}_M &= {D}_M J_M ,
\label{eq3.36} \\
J_M {\Gamma}_M &= {\Gamma}_M J_M .
\label{eq3.37}
\end{align}

In the Minkowskian signature, we shall define $\mathcal{J}_M$ by the same relation as Eq.(\ref{eq3.28}).  In this case, however, the charge conjugation is defined for Dirac spinors.  A Dirac spinor $\psi$ is composed  of two Weyl spinors.
%
%
The state $\Psi$ in (\ref{eq2.6}) and its charge conjugate state $\Psi^c$ in $\mathcal{H}_M$ is denoted by 
\eq
\Psi = \left(
\varphi_+ ,
\psi_{+\A} ,
F_ +,
\varphi_- ,
\psi_-^{\dot{\A}} ,
F_-
\right)^T,
\label{eq3.39}
\eqend
and
\eq
 \Psi^{c}  = \left( 
\varphi_-^\ast ,
\ovl{\psi}_{-\A} ,
F_-^\ast ,
\varphi_+^* ,
\ovl{\psi}_+^{\dot{\A}} ,
F_+^*
\right)^T.
\label{eq3.40}
\eqend
The charge conjugation matrix in Eq.(\ref{eq3.29}) is now given by
\eq
{C} = \left(
\begin{array}{@{\,}ccc|ccc@{\,}}
 & & & 1 & 0 & 0   \\
 & {\bf 0} & & 0 & \varepsilon_{\A\B} & 0   \\
 & & & 0 & 0 & 1  \\ \hline
 1 & 0 & 0 & & &  \\
 0 & \varepsilon^{\dot{\A}\dot{\B}} & 0 & & {\bf 0} & \\
 0 & 0 & 1 & & &
\end{array}
\right).
\label{eq3.41}
\eqend

The Z/2 grading  in the Minkowskian signature is defined by
\eq
\gamma_M = \begin{pmatrix}
-i & 0 \\
0 & i
\end{pmatrix},
\label{eq3.42}
\eqend
since the state $\Psi$ and its charge conjugate state $\Psi^c$ in $\mathcal{H}_M$ are given by Eq.(\ref{eq3.39}) and Eq.(\ref{eq3.40}) so that
\begin{align}
\gamma_M(\Psi_+) &= \gamma_M(\Psi_-^*) = \gamma_M(\Psi_+^c) = -i ,
\label{eq3.43} \\
\gamma_M(\Psi_-) &= \gamma_M(\Psi_+^*) = \gamma_M(\Psi_-^c) = i .
\label{eq3.44}
\end{align}
Although $\gamma_M$ and $C$ are anticommuting, $\mathcal{J}_M$ defined by Eq.(\ref{eq3.29}) commutes with $\gamma_M$ and satisfies the same relation as Eq.(\ref{eq3.31}).
As a result, the real structure  $J_M$ as well as  $D_M$ and $\Gamma_M$ in the Minkowskian signature satisfy the same relation as 
 Eq.(\ref{eq3.35})--Eq.(\ref{eq3.37}).
\section{Discussions and Conclusions} 

\ \ \
We have considered the supersymmetric extension of the spectral triple which define NCG in order to 
incorporate supersymmetry and obtained the supersymmetrically extended 
Dirac operator $\mathcal{D}_M$.  It acts on $\mathcal{H}_M$ that is also extended 
from the Hilbert space $C^\infty(M)$ to include not only spinor wave functions which represents matter fields 
but also their superpartners and auxiliary fields. Transferring from Minkowskian signature to Euclidean signature 
by the Wick rotation and applying the projection operator $P$ which transforms $\mathcal{H}_M$ to the subspace of 
spinors $L^2(M,S)$ embedded in $\mathcal{H}_M$, we can make the triple go back to the original spectral triple.    
The extended Dirac operator $\mathcal{D}_M$ inherits self-adjoint characteristics, compactness of its resolvent. 
$Z/2$ grading operator and antilinear operator which gives the real structure of $\mathcal{H}_M$ are also introduced.

But, $[\mathcal{D}_M,a]$ is not bounded for all $a\in \mathcal{A}_M$, because $\mathcal{D}_M$ contains second rank derivatives of 
space-time variables. So, we shall note that the triple does not define NCG extended from the original one. 
Our goal is not to construct supersymmetric new NCG, but to extend the theories of particles and interactions 
based on NCG to those incorporating supersymmetry. It includes the prescriptions to obtain the 
super Yang-Mills theory and minimum supersymmetric standard model  
such as supersymmetric version of spectral action principle. 
Indeed, as for the case without vector supermultiplets with gauge degrees of freedom,
we have shown that $\mathcal{D}_M$ 
provides the kinetic 
terms of matter particles and their superpartners in Eq.(\ref{Ikinetic}).  

In our next paper, we will introduce the "triple" in the finite space denoted by 
$(\mathcal{A}_F,\mathcal{H}_F,\mathcal{D}_F)$, where  
$\mathcal{H}_F$ is the space of labels which denotes quantum numbers of matter particles and  
$\mathcal{A}_F$ is the algebra represented on $\mathcal{H}_F$. 
We will show that the supersymmetrically extended Dirac operator $\mathcal{D}_F$ provides mass terms of 
matter particles and their superpartners. We will also in our next paper discuss the vector supermultiplet 
with gauge degrees of freedom. In the NCG theory without supersymmetry, gauge fields are introduced 
by the fluctuations of the Dirac operator $\Slash{D}$.  The matrix form of the Dirac operator 
$\Slash{D}$ and elements of algebra 
are given by
\begin{align}
\Slash{D} & = \begin{pmatrix}
0 & i\sigma^\mu\partial_\mu\\
i\bar{\sigma}^\mu\partial_\mu & 0
\end{pmatrix}, \\ 
a_j & = \begin{pmatrix}
\varphi_{2j} & 0\\
0 & \varphi_{2j}^\prime
\end{pmatrix}, \ \  
b_j =\begin{pmatrix}
\varphi_{1j} & 0\\
0 & \varphi_{1j}^\prime
\end{pmatrix},
\end{align}
where $\varphi_{1j},\varphi_{2j},\cdots$ are complex functions with internal degrees of freedom. 
The internal fluctuation to $\Slash{D}$ is given by
\begin{align}
\lefteqn{
\sum_j a_j[\Slash{D},b_j]}\nonumber \\
= & 
\begin{pmatrix}
0 & \sum_ji\sigma^\mu(\varphi_{2j} \partial_\mu\varphi_{1j}^\prime -\varphi_{2j}\varphi_{1j}\partial_\mu) 
\\
\sum_j i\bar{\sigma}^\mu(\varphi_{2j}^\prime\partial_\mu\varphi_{1j}-\varphi_{2j}^\prime\varphi_{1j}^\prime\partial_\mu) & 0
\end{pmatrix} \\
= &\begin{pmatrix}
0 & \sigma^\mu A_\mu\\
\bar{\sigma}^\mu A_\mu & 0
\end{pmatrix}.\ 
\end{align}
If the induced field $A_\mu$ is gauge field, it obeys the transformation law of the gauge field 
and satisfies the condition:  
\begin{equation}
A_\mu = A_\mu^\dagger \label{Ahermite}
\end{equation}
Authors choose a simple solution \cite{connes2} to the conditions which is given by 
\begin{align}
\varphi_{1j}^\prime & =\varphi_{1j},\ \varphi_{2j}^\prime=\varphi_{2j} \\
\varphi_{2j} & =\varphi_{1j}^\ast, \\
A_\mu & = \sum_j i\varphi_{1j}^\ast(\partial_\mu\varphi_{1j}). \label{AConnes}
\end{align}
%
In the supersymmetrically extended theory, 
vector supermultiplet which includes gauge field will be also introduced by 
the internal fluctuation due to elements of $\mathcal{A}_M$ to the 
Dirac operator $\mathcal{D}_M$ and will be given by the following form:
\begin{equation}
V_{i\bar{i};j\bar{j}}= \sum_a a_{i\bar{i};k\bar{k}}^\ast[\mathcal{D}_M,b]_{k\bar{k};j\bar{j}}, \label{fluctuationDM}
\end{equation}
where $a_{i\bar{i};j\bar{j}}$,$b_{i\bar{i};j\bar{j}} $ is an element of $\mathcal{A}_M$ in the form of (\ref{eq2.7}). 
As for the elements in (\ref{eq2.7}), however, 
since $u_{ij}$ is constructed by elements of chiral supermultiplet and yet $\bar{v}_{\bar{i}\bar{j}}$ is constructed by 
those of antichiral supermultiplet, $u_{ij}$ does not equal $\bar{v}_{\bar{i}\bar{j}}$ identically. 
So, we can not extend the solution (\ref{AConnes}) supersymmetrically. We must discover the solution in which each component of vector supermultiplet, gauge field,
gaugino and auxiliary field $D$ satisfies the adequate transformation laws. We will 
show that the fluctuation of the form of (\ref{fluctuationDM}) indeed produces 
vector supermultiplet with $U(N)$ gauge degrees of freedom and the heat kernel expansion of 
squared $\mathcal{D}_M$ modified by the fluctuation 
will give the action of super Yang-Mills theory\cite{SuYang,MSSM}. 

\end{document}